\documentclass[preprint,aps,ams,eqsecnum]{revtex4}

\begin{document}

\title{Decoherence in Quantum Systems{$^\dagger$}}

\author{R. F. O'Connell{$^\ddagger$}}

\affiliation{Department of Physics and Astronomy, Louisiana State
University,
Baton Rouge, LA  70803-4001}

\date{\today }

\begin{abstract}
We discuss various definitions of decoherence and how it can be measured.  We compare
and contrast decoherence in quantum systems with an infinite number of eigenstates (such
as the free particle and the oscillator) and spin systems.  In the former case, we point
out the essential difference between assuming "entanglement at all times" and
entanglement with the reservoir occuring at some initial time.  We also discuss optimum
calculational techniques in both arenas.
\\
\\
\\
\\
\\
\\
\\
{$^\dagger$}To be published in Proceedings of the 2004 IEEE NTC Quantum Device
Technology Workshop
\\
\\
\\{$^\ddagger$}Phone: (225)578-6848 \\
\noindent Fax: (225)578-5855 \\
\noindent Email: rfoc@phys.lsu.edu
\end{abstract}

\maketitle

\section{Introduction}

There is presently intense interest in the fundamentals of quantum theory and
applications.  In particular, the superposition principle (and related work on
Schr\"{o}dinger cats), entanglement, and the quantum-classical interface are at the
cutting-edge of topical research, especially in respect to their relevance to quantum
computing, quantum information processing, quantum teleportation and quantum
encryption.  Since superposition states are very sensitive to decoherence, reservoir
theory has attracted much recent interest.

Decoherence is the physical process which is responsible for the emergence of the
classical world from the quantum world
\cite{zeh,leggett,wigner,walls,joos,zurek,ford1,ford2}.  In order to understand the
phenomenon in depth, much attention has been devoted to the case of a free particle
interacting with a reservoir.  Most of these investigations have assumed that
entanglement with the reservoir occurs at some initial time, leading to the conclusion
that a dissipative environment is necessary to achieve decoherence.  However, we have
shown that in the case of "entanglement with the reservoir at all times", decoherence
can occur simply due to temperature effects without requiring dissipation
\cite{ford2,ford3,ford4}.  In addition, we pointed out that there are many definitions
of decoherence \cite{murakami} and we argued that the preferred definition involves
probabilities in coordinate space since they can be measured (in contrast to the
off-diagonal components of the density matrix, as considered often in the literature). 
Moreover, the fact that the free particle wave-packet spreads even in the absence of a
reservoir necessitates a careful consideration of how decoherence should be defined.

By contrast, an oscillator wave-packet does not spread in the absence of a reservoir so
that its investigation was clearly worthy of investigation \cite{ford5}.  This brought
to light another interesting feature viz. the fact that revivals of
coherence can occur.  A new dimension was added to the problem by consideration of an
external field
\cite{oconnell,zuo}, of interest because of the recent experiments of the Wineland
group \cite{myatt,turchette}.

Thus, in Secs. II and III, we present a detailed discussion of decoherence for both a
free particle and an oscillator, respectively, in an arbitrary reservoir, including
remarks on what we consider to be the optimal tools for such calculations.  Sec. IV is
devoted to a discussion of decoherence in spin systems and the significant differences
that arise compared to systems with an infinite number of eigenstates.  In Sec. V, we
summarize and discuss our conclusions.

\section{Decoherence for a Free Particle}

Decoherence refers to the destruction of a quantum inferference pattern and is relevant
to the many experiments that depend on achieving and maintaining entangled states. 
Examples of such efforts are in the areas of quantum teleportation, quantum information
and computation, entangled states, Schr\"{o}dinger cats, and the quantum-classical
interface.  Much of the discussion of decoherence has been in terms of a particle
moving in one dimension that is placed in an initial superposition state
(Schr\"{o}dinger "cat" state) corresponding to two widely separated Gaussian wave
packets, each with variance $\sigma^{2}$ and separated by a distance $d$.  We
consider the particle to be coupled to an arbitrary reservoir such that in the
distant past the complete system is in thermal equilibrium at temperature $T$.  For such
a state the probability distribution at time
$t$ can be shown to be of the form
\cite{ford2}

\begin{eqnarray}
P(x,t) &=&
\frac{1}{2(1+e^{-d^{2}/8\sigma^{2}})}\left\{P_{0}\left(x-\frac{d}{2},t\right)+P_{0}
\left(x+\frac{d}{2},t\right)\right. \nonumber \\
&&{}+\left.
2e^{-d^{2}/8\omega^{2}(t)}a(t)P_{0}(x,t)\cos\frac{[x(0),x(t)]xd}{4i\sigma^{2}w^{2}(t)}
\right\} \nonumber \\
&\equiv& P_{1}+P_{2}+2P_{I}\cos\theta(t) \label{dqs21}
\end{eqnarray}
where $P_{0}$ is the probability distribution for a single wavepacket, given by

\begin{equation}
P_{0}(x,t)=\frac{1}{\sqrt{2\pi w^{2}(t)}}\exp\left\{-\frac{x^{2}}{2w^{2}(t)}\right\}.
\label{dqs22}
\end{equation}
Here and in (\ref{dqs21}) $w^{2}(t)$ is the variance of a single wavepacket, which in
general is given by

\begin{equation}
w^{2}(t)=\sigma^{2}-\frac{[x(0),x(t)]^{2}}{4\sigma^{2}}+s(t), \label{dqs23}
\end{equation}
where $\sigma^{2}$ is the initial variance, $\left[x(0),x(t)\right]$ is the commutator,
and

\begin{equation}
s(t)=\left\langle \left\{x(t)-x(0)\right\}^{2}\right\rangle , \label{dqs24}
\end{equation}
is the mean square displacement.  The temperature dependence enters only in $s(t)$.
 Also,
$a(t)$, which can be defined as the ratio of the factor multiplying the cosine in the
interference term to twice the geometric mean of the first two terms \cite{ford2} is
given by the following exact general formula

\begin{equation}
a(t)=\exp\left\{-\frac{s(t)d^{2}}{8\sigma^{2}w^{2}(t)}\right\}. \label{dqs25}
\end{equation}

The question now arises as how to define decoherence.  It is often described as the
disappearance of the interference term with time.  However, this is not strictly
correct since the integrated probability of each of the three terms in (\ref{dqs21}) is
constant in time.  In fact, if we define the common normalization factor in
(\ref{dqs21}) as

\begin{equation}
N=\left[2\left(1+e^{-d^{2}/8\sigma^{2}}\right)\right]^{-1/2} \label{dqs26}
\end{equation}
we obtain

\begin{eqnarray}
\int^{\infty}_{-\infty}dxP(x,t) &=&
N^{2}\left\{1+1+2\exp\left(-\frac{d^{2}}{8\sigma^{2}}\right)\right\} \nonumber \\
&=& 2N^{2}\left\{1+\exp\left(-\frac{d^{2}}{8\sigma^{2}}\right)\right\} \nonumber \\
&=& 1. \label{dqs27}
\end{eqnarray}
It is clear that the integrated probability of each of the 3 terms is constant in
time.  Also, the integrated probability of the interference term is smaller than the
integrated probability of either of the direct terms by a factor
$\exp\left(-\frac{d^{2}}{8\sigma^{2}}\right)$ which, for macroscopic $d$ and
microscopic $\sigma$ is very small.  Thus, for example, when $(d/\sigma)=5$ we see that
$\exp\left(-\frac{d^{2}}{8\sigma^{2}}\right)\approx 4.4\times 10^{-2}$ which is a very
small number compared to unity.

In fact, the effect of decoherence is to cause the interference wave packet to spread
in time, with a concomitant decrease in its amplitude, consistent with the fact that its
integrated probability is constant in time.  Moreover, we note the ubiquituous $\cos$
term which is a feature of the interference contribution.  Thus, a measure of
decoherence is to focus on the interference term by calculating and measuring the time
dependence of $P_{I}$.  However, if decoherence is taken to mean the effects of a
reservoir (not necessarily dissipative) then this definition has a problem since it
given a non-zero value for the rate of decoherence in the absence of a reservoir.  The
problem stems from the fact that the wave packet for a free particle spreads naturally
in time, a standard result in quantum mechanics \cite{merzbacher}.  In fact, this result
immediately follows from (\ref{dqs23}) since, for a free particle without dissipation
and with zero temperature, $s(t)=0$ and $\left[x(0),x(t)\right]=i\hbar t/m$, so that

\begin{equation}
w^{2}(t)\rightarrow\sigma^{2}+\frac{\hbar^{2}t^{2}}{4m^{2}\sigma^{2}}. \label{dqs28}
\end{equation}
As a result, all of the three wave packets in (\ref{dqs21}) spread in time.  Thus, in
order to correct for this dynamical spreading, we were led to define \cite{ford2}

\begin{equation}
a(t)=\frac{P_{I}}{\left(P_{1}~P_{2}\right)^{1/2}}, \label{dqs29}
\end{equation}
as a measure of decoherence. It has the virtue of reducing to unity in the absence of a
reservoir \cite{ford4}.

In the case of Ohmic dissipation, high temperature $\left(kT>>\hbar\gamma\right)$,
small times $(\gamma t<<1)$ and
$d >>\lambda_{th},\sigma$, we obtained \cite{ford2,ford3} $s(t)=(kT/m)t^{2}$ and
$w^{2}(t)\approx \sigma^{2}$. Hence (\ref{dqs25}) reduces to

\begin{equation}
a(t)\rightarrow\exp\left\{-\left(\frac{t}{\tau_{d}}\right)^{2}\right\}, \label{dqs210}
\end{equation}
where

\begin{equation}
\tau_{d}=\frac{\sqrt{8}~\sigma^{2}}{\sqrt{kT/m}~d}, \label{dqs211}
\end{equation}
exhibiting "decoherence without dissipation".  Here $\gamma$ is the Ohmic decay rate,
and $\lambda_{th}=\hbar/\sqrt{mkT}$ is the deBroglie wavelength.  Hence

\begin{equation}
\lambda^{2}_{th}=\left(5\cdot 2\times 10^{-21}cm\right)^{2}\left(\frac{1gm}{m}\right)
\left(\frac{300K}{T}\right). \label{dqs212}
\end{equation}
For example, the choice $m=1gm,~d=1cm$ and $T=300K$, leads to the very large ratio
$d/\lambda_{th}=2\times 10^{20}$.  In addition, we note that 

\begin{equation}
\frac{kT}{\hbar\gamma}=\frac{T(K)}{\gamma\left(10^{11}s^{-1}\right)}. \label{dqs213}
\end{equation}
On the other hand, if we consider a low temperature environment $(kT<<\hbar\gamma
)$ \cite{ford6}, then using the quantum Langevin equation \cite{ford7} to evaluate
$w^{2}(t)$, we obtained

\begin{equation}
a(t)=\exp\left\{\left(\frac{t}{\tau_{0}}\right)^{2}\left[\log\frac{\zeta
t}{m}+\gamma_{E}-\frac{3}{2}\right]\right\},~~\tau <<t<<(\zeta /m)^{-1} \label{dqs214}
\end{equation}
with

\begin{equation}
\tau_{0}\equiv\frac{m\sigma^{2}}{d}\sqrt{\frac{8\pi}{\hbar\zeta}} \label{dqs215}
\end{equation}
and $\zeta=\gamma m$ and where $\gamma_{E}=0.577215665$ is Euler's constant. Thus, in
the low-temperature regime, decoherence requires a dissipative environment.

We emphasize again that all of the above results correspond to the case of the free
particle being entangled with the environment at all times.  Initially (or in the
distant past), the complete system is in equilibrium at temperature $T$ \cite{ford2}. 
At time $t=0$, say, a first measurement is carried out which prepares the system of two
widely separated wave packets.  Then, after a time $t$, a second measurement is carried
out which probes the system.

Most other investigations have assumed that the quantum system is initially at
temperature zero and uncoupled from the reservoir, which is at temperature $T$.  When
the quantum system is brought into contact with the reservoir at $t=0$, it takes a time
$\sim\gamma^{-1}$ in order for thermal equilibrium to be achieved, which is generally
much larger than the decoherence decay time so that decoherence occurs before thermal
equilibrium is achieved.  Such a calculation generally involves the use of density
matrix equations.  Thus, in order to make a careful comparison with our results
corresponding to "entanglement at all times" we solved the exact HPZ master equation
\cite{ford1} and obtained results for an arbitrary reservoir and arbitrary
temperatures.  In the particular case of high temperature, we obtained

\begin{equation}
a(t)\cong\exp\left\{-\frac{\zeta 
kTd^{2}t^{3}}{12m^{2}\sigma^{4}+3\hbar^{2}t^{2}}\right\},~~~~~t\ll m/\zeta\equiv\gamma
^{-1}. \label{dqs216}
\end{equation}
If we suppose that the slit width is negligibly small, we find
$a(t)\cong\exp\{-t/\tau_{d}\}$ where $\tau_{d}=\frac{3\hbar^{2}}{\zeta kTd^{2}}$.  This
is similar to the decoherence time that often appears in the literature.  But, as we
have mentioned above, this result corresponds to a particle in an initial state that is
effectively at temperature zero, which is suddenly coupled to a heat bath at high
temperature.  The result is therefore unphysical in the sense that the initial states
does not correspond to that envisioned when we speak of a system at temperature $T$. 
Thus, with this scenario, we get no decoherence, [i.e. $a(t)=1$] when $\gamma=0$.  On
the other hand, we found that we could repair this unphysical difficulty by choosing
the initial temperature to be the same as the reservoir, in which case the result given
in (\ref{dqs210}) is again obtained.

We conclude this section by commenting on the techniques used.  The "entanglement at
all times" calculation utilized quantum probability distributions (which are related to
Wigner distributions which are probabilities in quantum phase space
\cite{hillery,oconnell2,scully}) in conjunction with results obtained by use of the
stationery solution to the generalized quantum Langevin equation
\cite{ford7}.

For the case where the oscillator and reservoir are initially decoupled, we formulated
the problem in terms of the Langevin equation for the initial value problem.  In fact,
a master equation was not required but, in order to make contact with other
investigators, we actually derived the HPZ exact master equation from the Langevin
equation approach.  Moreoever, the Langevin approach enabled us to obtain explicit
results for the time-dependent coefficients in the HPZ equation.  Our strategy was
based on use of the Wigner distribution eventually leading to explicit and very general
results for the Wigner distribution for the reduced system of an oscillator in an
arbitrary state and in an arbitrary heat bath.  These results were then used to obtain
coordinate probabilities.

The Wigner distribution is, of course, the Fourrier transform of the density matrix
\cite{hillery}, so that results obtained by use of the former are equivalent to results
obtained by use of the latter.  However, Wigner functions are much easier to use in
practice because, as distinct from density matrices, they are not operators and they
are always real.  They describe, in essence, quantum phase space.  However, just as we
object to using the density matrix as a quantitative way to describe decoherence, we
also would have the same objections to the use of Wigner distributions for that purpose
\cite{murakami} but, instead, we regard them as a wonderful tool to calculate
probability distributions, which are measurable quantities.  A detailed review of the
use of Wigner distributions in this context appears in \cite{oconnell2}.

\section{Decoherence for an Oscillator}

A free oscillator wave packet does not spread in time (i.e. its width does not depend on
time) but it oscillates back and forth with a frequency $\omega$, where
$\omega$ is the oscillator frequency \cite{scully}.  In other words, its shape does not
change but the peak of the wave packet has the time dependence $x_{0}\cos\omega t$. 
Thus, in contrast to the free particle, there is no dynamical spreading for an
oscillator wave packet; it is a coherent state.

In the case of a Schr\"{o}dinger cat superposition of two coherent states separated at
a distance $d$ and in thermal equilibrium in a non-dissipative reservoir, we obtained a
result for $P(x,t)$ which is similar in structure to the result given in (\ref{dqs21}),
the difference being that the interference term, in common with the other terms, is
coherent in the sense that the wave packet shape does not change but, instead, it
oscillates in time and persists for all time.  Explicitly, the corresponding result for
$a(t)$ is \cite{ford8}

\begin{equation}
a(t)=\exp\left\{-\frac{m\omega d^{2}\cos^{2}\omega
t}{2\hbar\sinh\left(\frac{\hbar\omega}{kT}\right)}\right\}. \label{dqs31}
\end{equation}

It is interesting to note that for small times $(\omega t<<1)$ after the times for
which the attenuation factor has its maximum value of unity $(\cos\omega t=0)$, and if
we recall that the initial width $\sigma$ of the oscillator wave packet is given by
$\sigma^{2}=(\hbar /2m\omega )$, then for $\omega\rightarrow 0$ we find that $a(t)$
reduces to the result given in (\ref{dqs210}) and (\ref{dqs211}) for the free particle
case.  However, in contrast to the latter case, for later times, there is a revival of
$a(t)$ toward its maximum value.  We believe that this revivial of coherence in a
non-dissipative thermal reservoir is a common feature, the only exception being the
free particle.  For the case of $\gamma\neq 0$, we refer to \cite{ford1}. 

Finally, we remark that a non-random external field $f(t)$ does not cause decoherence,
in contrast to the case of a random $f(t)$ \cite{oconnell,zuo}, a result of relevance
for the analysis of the experimental results obtained by the Wineland group
\cite{myatt,turchette}.

\section{Decoherence in Spin Systems}

Whereas the study of decoherence for a free particle and an oscillator is very relevant
for the study of the classical-quantum correspondence, it is now generally believed
that spin systems (where the emphasis is on the spin and associated magnetic moment of
quantum particles) are more relevant to quantum computing, teleportation and
information processing.  There already exists extensive studies of such systems
\cite{wolf,awschalom,hogan} so we confine ourselves to some general remarks.

The main advantage of spin qubits are that they interact weakly with their
environment.  In general, as distinct from the study of free particle and oscillator
systems interacting with a reservoir, we do not have to worry to the same extent with
the question of when entanglement with the environment occurs since the system is
generally controlled by external fields.  Thus, density matrix techniques will
invariably constitute the tool of choice.  This brings up the question as to whether or
not there is an optimum way to solve these operator equations.  We already saw in Secs.
II and III that Wigner distribution functions proved to be a wonderful calculational
tool for the study of system with an infinite spectrum of states.  However, for spin
systems, there is also what we consider is a very useful and physically appealing tool
viz. the use of the spin polarization vector.  This has found application for particles
of arbitrary spin in a magnetic field but not interacting with a reservoir \cite{schiff}
but, strangely enough, it has not usually been adopted for the more challenging
situations where a reservoir is present. Ford and me actually used this technique in
order to calculate the spectrum of resonance fluorescence for a driven two-level atom
\cite{ford9}, instead of using the corresponding but more unwieldy master equation. 
Thus, in the Appendix, we use the spin polarization vector to study an even simpler
system.  This is the well-known two-level system (describing either a spin
$\frac{1}{2}$ system or a two-level atom) interacting with a reservoir and we feel
that, our calculation demonstrates the simplicity and physically appealing nature of
using the spin polarization vector.

\section{Discussion}

We have pointed out that, whereas it is easy to describe qualitatively what is meant by
decoherence, difference of opinions may arise as how best to define its signature in a
quantitative manner.  Different results may ensue depending on 

\noindent (a)~~whether one is dealing with coordinate, momentum or quantum phase space
probabilities (and we favor the former since they are measurable).

\noindent (b)~~how natural dynamical spreading of a wave packet in the case of a free
particle is separated from spreading due to environmental effects.

\noindent (c)~~whether one considers entanglement with a reservoir as existing for all
times or simply taking place at some initial time.

\noindent (d)~~whether external forces come into play.

\noindent (e)~~the system under consideration since quantum systems with an infinite
number of eigenstates (such as the free particle and the oscillator) generally exhibit
decoherence decay times much smaller than relaxation times whereas, for spin systems,
decoherence decay times are often comparable to relaxation times (which is the reason
they are preferred for applications).

In addition, we pointed out the great calculational advantages of using either Wigner
functions (for the free particle and the oscillator) or spin polarization vectors (for
spin systems).

\appendix

\section{}

{\bf{Solution of the master equation for a two-level system interacting with
a reservoir, using the spin polarization vector.}}

The master equation for the reduced density matrix $\rho$ may be written in the form

\begin{eqnarray}
\frac{d\rho}{dt} &=&
-\frac{\gamma}{2}\left(1+\bar{n}\right)\left\{\left[\sigma_{+},
\sigma_{-}\rho\right]+\left[\rho\sigma_{+}, \sigma_{-}\right]\right\}
\nonumber \\
&&{}-\frac{\gamma}{2}\bar{n}\left\{\left[\sigma_{-},\sigma_{+}\rho\right]+
\left[\rho\sigma_{-},\sigma_{+}\right]\right\} \nonumber \\
&=&
\frac{\gamma}{2}\left(1+\bar{n}\right)~\left(2\sigma_{-}\rho
\sigma_{+}-\sigma_{+}\sigma_{-}\rho-\rho\sigma_{+}\sigma_{-}\right) \nonumber
\\
&&{}+
\frac{\gamma}{2}\bar{n}\left(2\sigma_{+}\rho\sigma_{-}-
\sigma_{-}\sigma_{+}\rho - \rho\sigma_{-}\sigma_{+}\right), \label{app1}
\end{eqnarray}
where $\bar{n}$ is the thermal average boson number

\begin{equation}
\bar{n}=\left[\exp\left\{\hbar\omega /kT\right\}-1\right]^{-1}. \label{app2}
\end{equation}
Also

\begin{equation}
\sigma_{\pm}=\frac{1}{2}\left(\sigma_{x}\pm i\sigma_{y}\right), \label{app3}
\end{equation}
where the $\sigma_{i}$ are the usual Pauli spin matrices.

Now $\rho$ can be expanded in terms of the complete set $(I,~\sigma_{i})$
with real coefficients:

\begin{equation}
\rho=\frac{1}{2}\left(I+\vec{P}\cdot\vec{\sigma}\right), \label{app4}
\end{equation}
where $I$ is the identity matrix and $\vec{P}$ is the {\underline{polarization
vector}}.  This leads, in the usual manner to

\begin{equation}
\vec{P}=\langle\vec{\sigma}\rangle=Tr\left(\vec{\sigma}\rho\right),
\label{app5}
\end{equation}
i.e. the polarization vector is the ensemble average of the spin vector.  Thus

\begin{equation}
\frac{d\vec{P}}{dt}=Tr\left(\vec{\sigma}~\frac{d\rho}{dt}\right). \label{app6}
\end{equation}

Next, we note that

\newpage

\begin{eqnarray}
&Tr& \left\{ \vec{\sigma}\left(\sigma_{+}\sigma_{-}\rho
+\rho\sigma_{+}\sigma_{-}-2\sigma_{-}\rho\sigma_{+}\right)\right\} \nonumber
\\
&=&
\left\langle\vec{\sigma}\sigma_{+}\sigma_{-}+\sigma_{+}\sigma_{-}\vec{\sigma}-2\sigma
_{+}\vec{\sigma}\sigma_{-}\right\rangle \nonumber \\
&=& \left\langle\frac{1}{2}\vec{\sigma}\left(1+\sigma_{z}\right)+\frac{1}{2}
\left(1+\sigma_{z}\right)\vec{\sigma}+\left(1+\sigma_{z}\right)\hat{z}\right\rangle
\nonumber \\
&=& \left\langle\vec{\sigma}+\hat{z}+\hat{z}+\sigma_{z}\hat{z}\right\rangle
\nonumber
\\ &=& \vec{P}+\hat{z}\cdot \vec{P}\hat{z} + 2\hat{z}. \label{app7}
\end{eqnarray}
and

\begin{eqnarray}
&Tr& \left\{\vec{\sigma}\left(\sigma_{-}\sigma_{+}\rho +
\rho\sigma_{-}\sigma_{+}-2 \sigma_{+}\rho\sigma_{-}\right)\right\} \nonumber
\\
&=&
\left\langle\vec{\sigma}\sigma_{-}\sigma_{+}+\sigma_{-}\sigma_{+}\vec{\sigma}-2
\sigma_{-}\vec{\sigma}\sigma_{+}\right\rangle \nonumber \\
&=&
\left\langle\frac{1}{2}\vec{\sigma}\left(1-\sigma_{z}\right)+\frac{1}{2}\left(1-
\sigma_{z}\right)\vec{\sigma}-\left(1-\sigma_{z}\right)\hat{z}\right\rangle
\nonumber \\
&=& \left\langle\vec{\sigma}-2\hat{z}+\sigma_{z}\hat{z}\right\rangle \nonumber
\\ &=& \vec{P}+\hat{z}\cdot\vec{P}\hat{z}-2\hat{z}. \label{app8}
\end{eqnarray}
Hence, using (\ref{app1}), and (\ref{app6}) to (\ref{app8}),

\begin{eqnarray}
\frac{d\vec{P}}{dt} &=& -\frac{\gamma}{2}\left(1+\bar{n}\right)~\left(\vec{P}
+P_{z}\hat{z}+2\hat{z}\right) \nonumber \\
&&{}-\frac{\gamma}{2}\bar{n}\left(\vec{P}+P_{z}\hat{z}-2\hat{z}\right)
\nonumber \\
&=& -\frac{\gamma}{2}\left(2\bar{n}+1\right)~\left(\vec{P}+P_{z}\hat{z}\right)
-\gamma\hat{z}. \label{app9}
\end{eqnarray}
At equilibrium (where $P=P_{0}$)

\begin{equation}
\frac{d\rho}{dt}=0 \label{app10}
\end{equation}
so that

\begin{equation}
\frac{dP}{dt}=0. \label{app11}
\end{equation}
Hence, the right-side of (\ref{app9}) is zero, which implies

\begin{equation}
\vec{P}+P_{z}\hat{z}=-\frac{2\hat{z}}{2\bar{n}+1}. \label{app12}
\end{equation}
It is thus clear that, in {\underline{thermal equilibrium}} at temperature
$T$, $\vec{P}$ is along $\hat{z}$ i.e.

\begin{equation}
\vec{P}=P_{z}\hat{z}\equiv P_{0}\hat{z}. \label{app13}
\end{equation}
Thus, from (\ref{app12}), we obtain

\begin{equation}
P_{0}=-\frac{1}{2\bar{n}+1}=-\tanh\left(\frac{\hbar\omega}{2kT}\right).
\label{app14}
\end{equation}
For $T=0(\bar{n}=0)$, we see that $P_{0}=-1$ corresponding to all the spins
being down.  On the other hand, for $T\rightarrow\infty
(\bar{n}\rightarrow\infty )$, we obtain $P_{0}\rightarrow 0$, corresponding
to the equal number of up and down spins.  Hence, (\ref{app9}) becomes

\begin{equation}
\frac{d\vec{P}}{dt}=\frac{\gamma}{2P_{0}}\left(\vec{P}+P_{z}\hat{z}\right)
-\gamma\hat{z}. \label{app15}
\end{equation}
Hence

\begin{eqnarray}
\frac{dP_{z}}{dt} &=& \frac{\gamma}{P_{0}}~P_{z}-\gamma \nonumber \\
&=& \frac{P_{0}-P_{z}}{T_{1}}, \label{app16}
\end{eqnarray}
where

\begin{equation}
\frac{1}{T_{1}}\equiv -\frac{\gamma}{P_{0}}=\gamma\left(2\bar{n}+1\right)=\gamma
\coth\left(\frac{\hbar\omega}{2kT}\right).
\label{app17}
\end{equation}
Thus, for $T\rightarrow 0$, $T_{1}\rightarrow\gamma^{-1}$ whereas for
$T\rightarrow\infty$, $T_{1}\rightarrow 0$. The solution of (\ref{app16}) is

\begin{equation}
P_{z}=P_{0}\left\{1-\exp\left(-\frac{t}{T_{1}}\right)\right\}. \label{app18}
\end{equation}
In addition, (\ref{app15}) may be written as

\begin{equation}
\frac{d\vec{P}}{dt}=-\frac{1}{2T_{1}}\left(\vec{P}+P_{z}\hat{z}\right)-\gamma\hat{z}.
\label{app19}
\end{equation}
Thus, defining $P_{\perp}$ to be the component of $\vec{P}$ perpendicular to
$\hat{z}$, we obtain

\begin{equation}
\frac{dP_{\perp}}{dt}=-\frac{P_{\perp}}{2T_{1}} \label{app20}
\end{equation}
whose solution is

\begin{equation}
P_{\perp}=P_{\perp}(0)\exp\left(-\frac{t}{T_{2}}\right), \label{app21}
\end{equation}
where

\begin{equation}
T_{2}\equiv 2T_{1}. \label{app22}
\end{equation}
Since the magnetic moment $\vec{M}$ of a nuclear ion with spin
$\vec{S}=\frac{\hbar}{2}\bar{\sigma}$ is given by

\begin{eqnarray}
\vec{M} &=&
g_{n}\frac{e}{2mc}\vec{S}=\frac{e\hbar}{2mc}~\frac{g_{n}}{2}\vec{\sigma}
\nonumber \\ &=& \mu_{0}\frac{g_{n}}{2}\vec{\sigma}, \label{app23}
\end{eqnarray}
where $m$ is the nuclear mass, $e$ is its charge, $g_{n}$ is the nuclear
ion $g$ factor and

\begin{equation}
\mu_{0}=\frac{e\hbar}{2mc}, \label{app24}
\end{equation}
is the nuclear Bohr magneton, it follows that

\begin{equation}
\left\langle \vec{M}\right\rangle=-\mu_{0}\frac{g_{n}}{2}\left\langle
\vec{\sigma}\right\rangle=-\mu_{0}\frac{g_{n}}{2}\vec{P}. \label{app25}
\end{equation}
Hence, using (\ref{app18}) and (\ref{app21}), we obtain

\begin{equation}
\left\langle
M_{z}\right\rangle =M_{o}\left\{1-\exp\left(-\frac{t}{T_{1}}\right)\right\},
\label{app26}
\end{equation}
and

\begin{equation}
\left\langle M_{\perp}\right\rangle
=M_{\perp}(0)\exp\left(-\frac{t}{T_{2}}\right). \label{app27}
\end{equation}
We note that $T_{1}$ is the so-called spin-lattice (longitudinal) relaxation time which
describes the approach to thermal equilibrium whereas $T_{2}$ is the dephasing
(transverse) relaxation time, describing a system of spin $\frac{1}{2}$ particles in a
magnetic field.

It is also of interest to re-write (\ref{app4}) in the form

\begin{equation}
\rho=\frac{1}{2}\left\{1+\left(\sigma_{+}P_{-}+\sigma_{-}P_{+}\right)
+\sigma_{z}P_{z}\right\}, \label{app28}
\end{equation}
where

\begin{equation}
P_{\pm}=\left(P_{x}\pm iP_{y}\right), \label{app29}
\end{equation}
so that, as is also obvious from (\ref{app19}) to (\ref{app21}),

\begin{equation}
\frac{dP_{\pm}}{dt}=-\frac{P_{\pm}}{2T_{1}}, \label{app30}
\end{equation}
and

\begin{equation}
P_{\pm}=P_{\pm}(0)\exp\left(-\frac{t}{T_{2}}\right).
\label{app31}
\end{equation}

Moreover, if we write

\begin{equation}
\rho=\left(
\begin{array}{ll}
\rho_{++}  &  \rho_{+-} \\
\rho_{-+}  &  \rho_{--}
\end{array}
\right), \label{app32}
\end{equation}
and use the fact that

\begin{equation}
\sigma_{+}=
\left(
\begin{array}{ll}
0 & 1 \\
0 & 0
\end{array}
\right),~~~~~\sigma_{-}=\left(
\begin{array}{ll}
0 & 0 \\
1 & 0
\end{array}
\right), \label{app33}
\end{equation}
it readily follows from (\ref{app28}) that

\begin{equation}
\rho_{++}=\frac{1}{2}\left(1+P_{z}\right) \label{app34}
\end{equation}

\begin{equation}
\rho_{--}=\frac{1}{2}\left(1-P_{z}\right) \label{app35}
\end{equation}

\begin{equation}
\rho_{+-}=\frac{1}{2}\left(1+P_{-}\right), \label{app36}
\end{equation}
and

\begin{equation}
\rho_{-+}=\frac{1}{2}\left(1+P_{+}\right). \label{app37}
\end{equation}
Thus, using (\ref{app18}) and (\ref{app31}), it is clear that we obtain the
familiar result that the rate of decay of the diagonal elements of the
density matrix $\left(T_{1}^{-1}\right)$ is twice that of the rate of decay of
the off-diagonal elements $\left(T_{2}^{-1}\right)$.  However, we suggest
that it is physically more appealing to say that we conclude [from
(\ref{app18}) and either (\ref{app21}) or (\ref{app31})] that the
expectation value of the spin component in the $z$ direction (direction of
the magnetic field) decays faster than the corresponding values perpindicular
to the $z$ direction.

\acknowledgements
The author is pleased to acknowledge that all of the essential results described above
were derived in collaboration with Professor G. W. Ford.

\end{document}